**Voltage-Gated Modulation of Domain Wall Velocity in an Ultrathin Metallic Ferromagnet**

Uwe Bauer, Satoru Emori and Geoffrey S. D. Beach[a]

Department of Materials Science and Engineering, Massachusetts Institute of Technology, Cambridge, Massachusetts 02139, USA

**Abstract**

The influence of gate voltage, temperature and magnetic field on domain wall (DW) creep dynamics is investigated in Pt/Co/GdOx films with perpendicular magnetic anisotropy and imaged by a scanning magneto-optical Kerr effect technique. The DW creep velocity can be controlled by an electric field applied to the Co/GdOx interface via a linear modulation of the activation energy barrier with gate voltage. At low speeds, the DW velocity can be changed significantly by a gate voltage, but the effect is diminished as the DW velocity increases, which limits electric field control of fast DW motion.

[a] Author to whom correspondence should be addressed. Electronic mail: gbeach@mit.edu

Magnetic domain walls (DWs) can encode information for a variety of spintronic memory and logic devices.[1-3] Achieving efficient electrical control of DWs is essential to realizing high-performance solid-state operation of such devices. Much work has been devoted to exploiting spin-polarized electric currents to manipulate magnetic DWs via spin-transfer torques.[4,5] While significantly more efficient than conventional magnetic fields, this mechanism is still dissipative and often suffers from a high critical current for DW displacement. Direct voltage-gated control of DW motion is therefore highly desirable, and has been achieved using strain-coupled magnetostrictive/piezoelectric composites,[6-10] but these materials pose challenges to CMOS integration. Recently, it has been shown that magnetic anisotropy at the interface between an ultrathin metallic ferromagnet and an oxide dielectric can be modulated by an applied electric field.[11-16] This effect provides a mechanism to achieve direct voltage control of DW propagation in transition metal ferromagnets.[17,18] However, understanding the efficiency and limitations of electric field control of DW dynamics is critical for assessing the potential of voltage-gated DW devices.

In this letter we examine DW dynamics in an ultrathin Co film under the influence of an electric field applied across a gate dielectric. We have recently demonstrated that a gate voltage can enhance or retard DW propagation in Pt/Co/GdOx films, leading to a linear modulation of the propagation field for DW creep motion.[17] Here we present direct space- and time-resolved measurements of DW dynamics spanning more than four decades in velocity using a high-resolution scanning magneto-optical Kerr effect (MOKE) polarimeter. By measuring the velocity scaling with temperature, driving field, and gate voltage, we verify domain expansion via thermally-activated creep dynamics and derive an experimentally-motivated expression to incorporate magnetoelectric effects into the creep equation of motion. We show that an electric field linearly modulates the activation energy barrier $E_A$ that governs DW creep, leading to an exponential dependence of DW velocity on gate voltage. As a consequence, significant voltage-induced velocity enhancement can be achieved in the low-velocity regime, but the efficiency of voltage control is diminished at high velocities where $E_A$ is correspondingly small, which limits voltage control of fast DW motion.

Experiments were performed on perpendicularly-magnetized Ta(4 nm)/Pt(3 nm)/Co(1 nm)/GdOx(40 nm) films grown at room temperature on thermally-oxidized Si(100). Here, the GdOx layer serves as a high-k gate dielectric[19] while simultaneously promoting PMA in the Co film due to Co-O interfacial hybridization.[20-22] The metal layers were grown by dc magnetron sputtering at 3 mTorr of Ar with a background pressure of ~1 x $10^{-7}$ Torr. The GdOx layer was reactively sputtered at an oxygen partial pressure of ~ 5x$10^{-5}$ Torr. After breaking vacuum, 100 μm diameter Ta(1 nm)/Au(5 nm) gate electrodes were deposited through a shadow mask. Magnetic properties were characterized using MOKE and vibrating sample magnetometry (VSM). VSM measurements yielded a saturation magnetization of ~1200 emu/(cm$^3$ of Co), suggesting minimal Co oxidation during growth of the GdOx overlayer. The films exhibit square out-of-plane hysteresis loops with a coercivity of ~250 Oe, and have an in-plane saturation field of ~6 kOe, indicating strong PMA.

DW motion in the films was studied using the technique introduced in Ref. 17 and described schematically in Fig. 1(a). A 25 μm diameter blunt W microprobe was used to generate an artificial DW nucleation site through application of a local mechanical stress. After initially saturating the film magnetization, a reversed magnetic field step $H$ was applied using an electromagnet with a ~300 μs risetime. The driving field nucleates a reversed domain underneath the W probe tip, which then expands radially across the film. Magnetization reversal was detected via the polar MOKE signal using a ~3 μm diameter focused laser spot positioned by a high-resolution scanning stage. Exemplary time-resolved MOKE signal transients are shown in Fig. 1(b) for several positions of increasing distance from the W probe tip. Each transient corresponds to the averaged signal acquired from 50 reversal cycles to account for stochasticity, and hence represents the integrated probability distribution of switching times at a given location.[23] The mean reversal time ($t_{1/2}$), taken as the time at which the probability of magnetization switching is 50 %, increases linearly with increasing distance from the nucleation site as expected for DW propagation. The sharp transitions in these averaged transients indicate that the DW motion is highly repeatable from cycle to cycle.

By acquiring MOKE signal transients across a two dimensional grid in the vicinity of the artificial nucleation site, the average DW trajectory can be investigated and imaged with μm spatial and μs time resolution. Figures 2(a)-(c) show maps of the normalized polar MOKE signal, corresponding to the perpendicular magnetization component $M_z$, at several times $t$ after application of a field step $H$ = 170 Oe. Figure 2(d) shows a corresponding contour map of the mean reversal time $t_{1/2}$ over the same 300 μm x 300 μm grid area. Here each pixel corresponds to the averaged measurements of 50 reversal cycles. These images clearly show that a reversed domain nucleates beneath the W probe tip and expands radially and isotropically with time. Natural nucleation sites elsewhere in the film are sufficiently distant such that no other DWs reach the probed region under the conditions examined. The remarkably sharp and azimuthally-uniform (averaged) DW propagation front suggests smooth DW motion via thermally-activated creep through a fine-scale disorder potential. While DW creep is stochastic on the scale of the disorder potential, the DW propagation is deterministic at the micrometer-scale resolution of the present experiments.

The dotted circle in Fig. 2(a) outlines the position of a Ta/Au electrode located ~50 μm from the artificial nucleation site. In the absence of a gate voltage, the propagating DW passes unimpeded underneath the electrode. In Figs. 2(e)-(g) a mechanically-compliant 15 µm diameter BeCu probe tip was gently landed on the electrode to apply a gate voltage $V_g$ without creating an additional nucleation site. In the experiments presented here, $V_g$ was limited to the range of ± 8 V to minimize charge trapping effects and associated irreversibility in the gate oxide.[16,17] Figures 2(e)-(g) show maps of $M_z$ for $V_g = 8$ V at exactly the same $t$ as the $V_g = 0$ maps of Figs. 2(a)-(c). Comparison reveals that at $V_g = 8$ V, domain expansion is impeded underneath the electrode whereas it occurs unimpeded outside of the electrode area. The mean switching time $t_{1/2}$ versus position, plotted in Fig. 2(h), highlights the effect of $V_g$ on DW propagation under the gate electrode. At $V_g = 0$, $t_{1/2}$ increases linearly as a function of distance from the nucleation site with the same slope inside and outside of the electrode area. However, for $V_g = 8$ V the slope in the electrode area increases whereas it decreases for $V_g = -8$ V, which indicates that beneath the electrode the DW velocity $v$ increases for $V_g < 0$ and decreases for $V_g > 0$.

In order to better understand and quantify the observed DW dynamics and the role of $V_g$ on these dynamics, we have measured DW velocity $v$ as a function of $H$, $V_g$, and substrate temperature $T$. The average DW velocity $v$ was determined from a linear fit of $t_{1/2}$ versus position from typically 9 measurements of $t_{1/2}$ over a distance of 80 µm across the electrode, on a line extending radially from the artificial nucleation site. In the creep regime of DW dynamics, DW motion is thermally activated and follows an Arrhenius-type relation $v \propto \exp(-E_a(H)/kT)$. Here, $k$ is the Boltzmann constant, $T$ the temperature and $E_a$ the activation energy barrier for DW propagation. $E_a$ is expected to follow a scaling law $E_a \propto (H_{crit}/H)^\mu$ where $H_{crit}$ is a characteristic depinning field, and the scaling exponent $\mu = 1/4$ for a one-dimensional elastic line in a two-dimensional disorder potential.[24]

We first measured $v(T)$ at $H = 170$ Oe and $V_g = 0$ V, spanning a temperature range from 25°C to 55 °C, controlled using a thermoelectric module stable to within $\pm 0.1$ °C. Within this range of $T$, the velocity spans nearly an order of magnitude. As seen in Fig. 3(a), $ln(v)$ scales linearly with $T^{-1}$, confirming Arrhenius-like velocity scaling.[25] The slope corresponds to a thermal activation energy barrier $E_a = (0.54 \pm 0.01)$ eV, or ~20 $kT$. In Fig. 3(b), we plot the dependence of $v$ on $H$ over a driving field range from 70 Oe to 230 Oe, corresponding to a change in $v$ by almost four decades. Within this range, the dependence of $v$ on $H$ is well-described via a scaling relation $\ln(v) \propto H^{-1/4}$, consistent with two-dimensional DW creep dynamics.[24]

Having established that DW motion is indeed in the thermally-activated creep regime, we proceed to examine the dependence of $v$ on $V_g$. The depinning field $H_{\text{crit}}$ is a function of the DW elastic energy density $\varepsilon_{el} = 4(AK_u)^{1/2}$ and the DW width $\delta = (A/K_u)^{1/2}$, where $A$ is the exchange constant and $K_u$ is the uniaxial anisotropy constant.[24] Therefore, $H_{\text{crit}}$ is expected to scale with $K_u$ and voltage induced changes to $K_u$ are expected to result in a direct modification of $E_a$ and therefore of $v$. In Fig. 3(c) the dependence of $v$ on $V_g$ is shown at $H = 170$ Oe and $T = 35$ °C. The DW velocity $v$ can be modified by a factor of two in a bias range of $V_g = -8$ to $+8$ V. Due to limited charge trapping likely at the Co/GdOx interface,[17] we find that the $v$ versus $V_g$ curves often show small hysteresis. Therefore, we plot $v$ averaged over several forward and reverse $V_g$ cycles. From Fig. 3(c) it can be seen that $\ln(v)$ and therefore $E_a$ scale approximately linearly with $V_g$. To accurately access the scaling exponent would require measurements over several decades in $v$ which is experimentally inaccessible. However, in the investigated bias range the scaling relation for $v$ can be expanded to include voltage induced effects:

$$v(V_g, H, T) \propto exp\left(-\left(1 + \frac{\alpha}{d_{Ox}} V_g\right) E_{a,0V}(H)/kT\right) \quad (1)$$

Here, $d_{Ox}$ is the GdOx thickness, $\alpha$ is a magnetoelectric proportionality factor and $E_{a,0V}$ is the activation energy barrier at $V_g = 0$ V. From the linear fit in Fig. 3(c), the previously determined $E_a$ at $V_g = 0$ V, and taking $d_{Ox} = 40$ nm, we find $\alpha = (8.1 \pm 0.5) \times 10^{-2}$ nm/V.

Given the full functional dependence of $v$ on $H$, $T$ and $V_g$, it is now possible to investigate in more detail the efficiency of voltage control of DW motion across the parameter space. The form of Eq. 1 suggests that the relative influence of $V_g$ on $v$ depends on the magnitude of $E_{a,0V}$. The relative DW velocity enhancement, defined here as the ratio between DW velocity at $V_g = -4$ V ($v_{-4V}$) and DW velocity at $V_g = 0$ V ($v_{0V}$) was determined at $T = 35$ °C in a driving field range of $H = 100$ to 230 Oe. As $H$ increases, $v_{0V}$ increases exponentially but the relative enhancement $v_{-4V}/v_{0V}$ decreases (see Fig. 4(a)). At $v_{0V} \approx 2 \times 10^{-4}$ m/s, corresponding to $H = 100$ Oe, we find a voltage-induced velocity enhancement of ~34 % at $V_g = -4$ V. However, this same $V_g$, leads to only a ~14% velocity enhancement at $v_{0V} \approx 1 \times 10^{-1}$ m/s, corresponding to $H = 230$ Oe. Therefore, the efficiency of voltage induced control of DW motion decreases with increasing DW velocity due to the correspondingly lower activation energy barrier.

This behavior follows directly from Eq. (1): an increase in $H$ results in a decrease of $E_{a,0V}$ which then reduces the effect of $V_g$ on the creep velocity. The ratio $v_{Vg}/v_{0V}$ can be expressed as

$$\ln(v_{Vg}/v_{0V}) = -\alpha V_g E_{a,0V}/d_{Ox} kT \qquad (2)$$

Therefore, $\ln(v_{-4V}/v_{0V})$ is expected to decrease linearly with $E_{a,0V}$ at a slope proportional to the magnetoelectric coefficient $\alpha$. We have determined $E_{a,0V}$ through direct Arrhenius analysis of $\ln(v)$ versus $T^{-1}$ curves, such as shown in Fig. 3(a), at driving fields $H = 110$, 130, 150, 170 and 200 Oe. We find that within the measurement accuracy $\alpha$ is independent of $H$. By plotting the velocity enhancement $v_{-4V}/v_{0V}$ directly as a function of $E_{a,0V}$ (see Fig. 4(b)) we find that indeed, $v_{-4V}/v_{0V}$ decreases with decreasing $E_{a,0V}$. From the linear fit in Fig. 4(b) $\alpha$ can be determined to $\alpha = (9.4 \pm 2.0) \times 10^{-2}$ nm/V, which agrees well with the value determined above within the measurement accuracy.

In summary, we demonstrate direct electric field control of DW creep dynamics in Pt/Co/GdOx films. DW motion was investigated as a function of gate voltage, temperature and magnetic field and imaged by a scanning MOKE technique. We find that DW velocity can be controlled by an electric field at the Co/GdOx interface and that the voltage-induced effects can be described by a linear modulation of the activation energy barrier via voltage-induced changes to the interfacial anisotropy. At low DW speeds corresponding to a large activation energy barrier, DW velocity can be significantly modified by a gate voltage. However, with increasing DW velocity the efficiency of voltage control is diminished due to the simultaneous decrease in activation energy, which limits voltage control of DW motion at high velocities.

**Acknowledgement**

This work was supported under NSF-ECCS -1128439. S.E. acknowledges financial support by the NSF Graduate Research Fellowship Program. Technical support from David Bono and Mike Tarkanian is gratefully acknowledged.

**Figure Captions**

FIG. 1. (color online) (a) Schematic illustration of experimental setup including gated Pt/Co/GdOx structure, (1) BeCu microprobe to apply gate voltage ($V_g$), (2) W microprobe to create artificial DW nucleation site and focused MOKE laser beam to map out (x, y) magnetic domain expansion. White arrows illustrate local orientation of magnetization vector during expansion of domain from artificial nucleation site. (b) Normalized MOKE signal transients measured at different distances *s* between MOKE probing spot and mechanically created nucleation site. Dashed line corresponds to 50% probability of magnetization switching.

FIG. 2. (color online) MOKE maps of perpendicular component of magnetization vector ($M_z$), showing domain expansion from artificial nucleation site at $t$ = 6.3 ms ((a), (e)), 11.9 ms ((b), (f)) and 14.7 ms ((c), (g)) after application of magnetic field step, with no gate voltage applied ((a)-(c)) and at $V_g$ = 8 V ((e)-(g)). (d) Contour map showing mean switching time ($t_{1/2}$) around nucleation site with no gate voltage applied. (h) Line scan of $t_{1/2}$ across gate electrode at $V_g$ = -8, 0 and + 8 V for another device with larger distance between electrode and nucleation site. Black triangular area on right side of (a)-(g) and on left side of (e)-(g) corresponds to W and BeCu microprobe tip, respectively. Dashed black line in (a), (e) outlines gate electrode. All measurements at $H$ = 170 Oe and $T$ = 35 °C.

FIG. 3. (color online) DW velocity $v$ as a function of (a) Temperature ($T$) at $H = 170$ Oe and $V_g = 0$ V, (b) magnetic field ($H$) at $T = 35$ °C and $V_g = 0$ V and (c) gate voltage ($V_g$) at $H = 170$ Oe and $T = 35$ °C. Red lines are linear fits of data. Error bars in (a) and (b) are smaller than plot symbols.

FIG. 4. (color online) (a) Ratio between DW velocity at $V_g = -4$ V ($v_{-4V}$) and at 0 V ($v_{0V}$) as a function of DW velocity $v_{0V}$. (b) Subset of same data plotted as a function of the activation energy barrier at $V_g = 0$ V ($E_{a,0V}$). Red lines are linear fits of data.

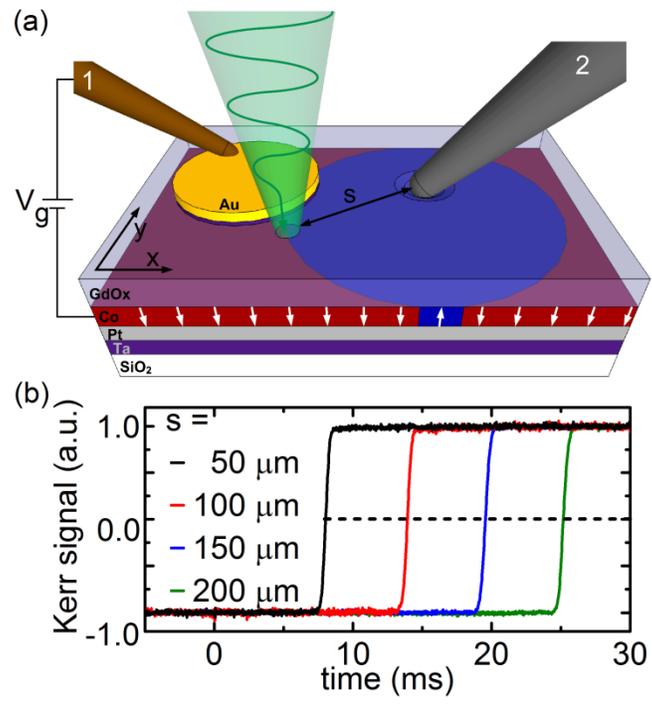

**Figure 1**

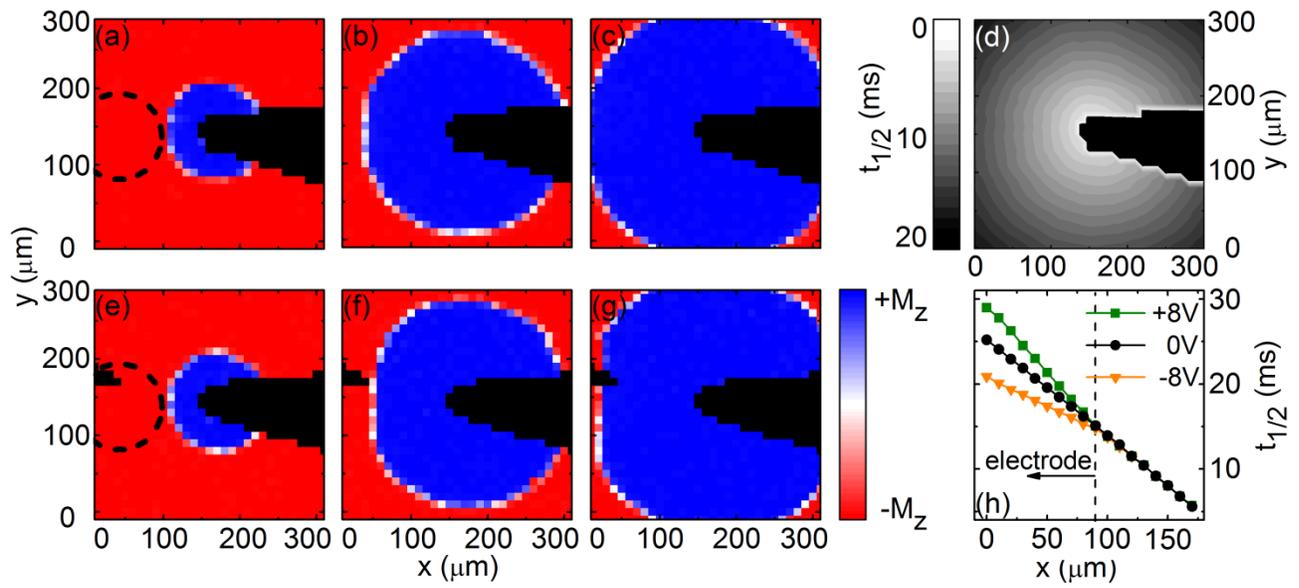

**Figure 2**

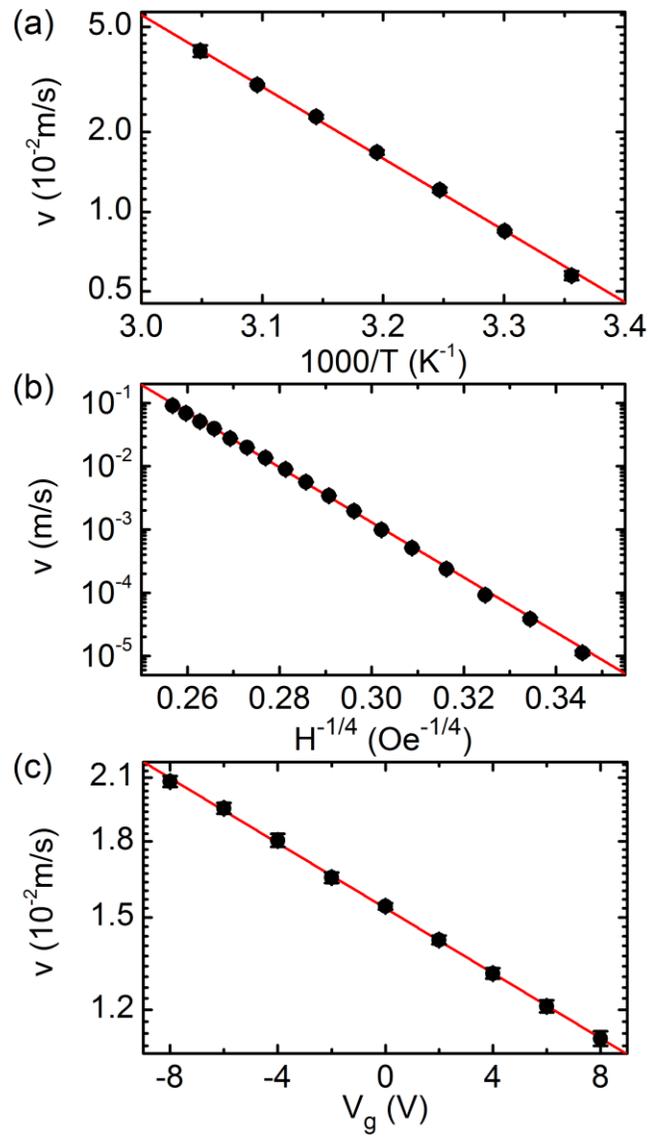

**Figure 3**

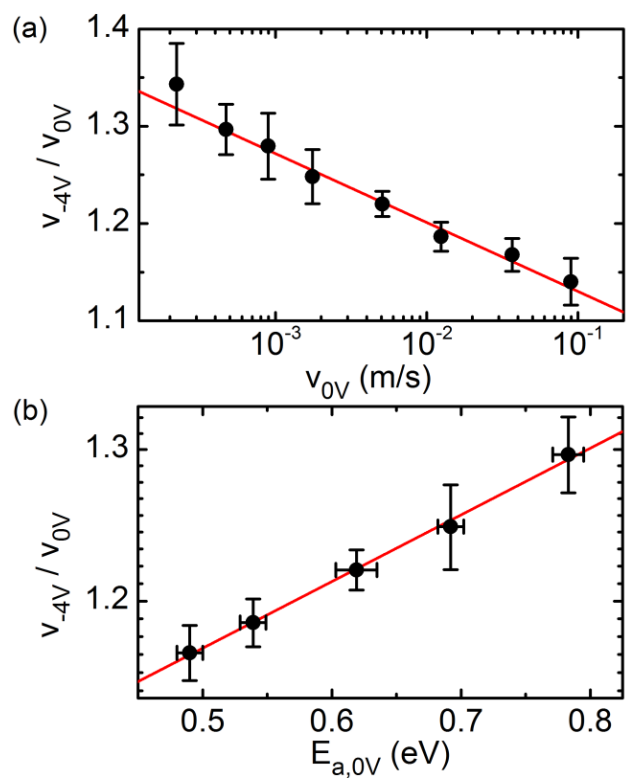

**Figure 4**